\documentclass[]{osa-article}
\pdfoutput=1
\journal{osajournal}
\usepackage{amsmath,amsfonts,amssymb}
\usepackage{graphicx}

\usepackage[separate-uncertainty=true]{siunitx}
\usepackage{subcaption}

\let\oldeqref\eqref
\renewcommand{\eqref}[1]{Eq.~\oldeqref{#1}}

\begin{document} 

\title{Probing multipulse laser ablation by means of self-mixing interferometry}

\author{Simone Donadello\authormark{1,*}, Ali G\"{o}khan Demir\authormark{1}, Barbara Previtali\authormark{1}}

\address{\authormark{1}Department of Mechanical Engineering, Politecnico di Milano, Via La Masa 1, 20156 Milan, Italy}
\email{\authormark{*}Corresponding author: simone.donadello@polimi.it}

\begin{abstract}
	In this work, self-mixing interferometry (SMI) is implemented inline to a laser microdrilling system to monitor the machining process by probing the ablation-induced plume. An analytical model based on the Sedov--Taylor blast wave equation is developed for the expansion of the process plume under multiple-pulse laser percussion drilling conditions. Signals were acquired during laser microdrilling of blind holes on stainless steel, copper alloy, pure titanium, and titanium nitride ceramic coating. The maximum optical path difference was measured from the signals to estimate the refractive index changes. An amplitude coefficient was derived by fitting the analytical model to the measured optical path differences. The morphology of the drilled holes was investigated in terms of maximum hole depth and dross height. The results indicate that the SMI signal rises when the ablation process is dominated by vaporization, changing the refractive index of the processing zone significantly. Such ablation conditions correspond to limited formation of dross. The results imply that SMI can be used as a nonintrusive tool in laser micromachining applications for monitoring the process quality in an indirect way.
\end{abstract}

\section{Introduction}
High-power lasers are widely used for machining materials when quality and fast manufacturing processes are required, allowing micrometric precision in many applications such as cutting, welding, hole drilling, and surface texturing \cite{dubey_laser_2008}. The radiation-matter interaction leads to modifications in the absorbing material, such as melting, evaporation, or sublimation, depending on the material characteristics and on the laser parameters. The usage of pulsed lasers allows very high energy densities to be obtained, concentrated on small areas and in short time intervals, hence optimizing the material ablation driven by the high peak power with a limited heat-affected zone \cite{chichkov_femtosecond_1996}. As a result of the laser ablation process, a material plume, composed of vapors, plasma, particles, or droplets, is ejected from the workpiece \cite{amoruso_characterization_1999}. The laser pulse duration is an important parameter for determining the ablation regime: for ultrashort pulses, i.e., with femtosecond and picosecond duration, the ablation is dominated by cold interaction with an effective transition from solid to vapor, while for short, nanosecond, pulses the material removal results from the combination of melting and evaporation phenomena \cite{zhigilei_dynamics_2003,povarnitsyn_material_2007,leitz_metal_2011}.

The ablation plume carries information about the ablation rate and efficiency, and its observation can be exploited to monitor the machining process, although the interpretation of its dynamics is not trivial \cite{wood_dynamics_1997,wood_dynamics_1998-1}. The problem of the plume evolution is typically approached by describing its expansion in analogy of the Sedov--Taylor blast wave theory, with a shock wavefront expanding from a point-like instantaneous explosion. Accordingly, the plume generation has been extensively studied with experiments in the case of ablation induced by a single laser pulse, in both the regimes of ultrashort \cite{konig_plasma_2005,zeng_experimental_2005,henley_dynamics_2005} and short pulses \cite{grun_observation_1991,gupta_direct_1991,geohegan_fast_1992,callies_time-resolved_1995,harilal_internal_2003,sharma_plume_2005,demir_investigation_2015}. However, it should be noted that many practical applications, such as percussion microdrilling, require multiple laser pulses with high repetition rates, where the plume dynamics becomes turbulent and it cannot be described directly as a single blast wave \cite{breitling_material-vapor_2003,breitling_fundamental_2004}.

Several optical techniques, including photography, interferometry, holography, or absorption imaging, represent convenient approaches for the observation of the fast-evolving phenomena of ablation plume expansion, allowing non-contact measurements with high temporal and spatial resolutions \cite{dyer_dynamics_1990,amer_comparison_2010,sangines_two-color_2011,gojani_identification_2013,choudhury_time_2016}. However, such methods are typically based on complex setups and require post-process analysis. On the contrary, the inline monitoring for the quality control in production environments should be based on simple and robust techniques with non-intrusive setups. Within this context, self-mixing interferometry (SMI) is gaining interest in mechanical engineering, since it allows for effective and cheap solutions in many kind of applications \cite{donati_laser_1995,giuliani_laser_2002,taimre_laser_2015,donati_overview_2018}, such as displacement, velocity, or vibration sensors, with sub-micrometer resolutions. The SMI technique has been successfully employed also for monitoring the laser microdrilling process. Several works reported the usage of diode lasers in a self-mixing configuration for measuring the displacement of the ablation front related to the laser ablation rate \cite{mezzapesa_high-resolution_2011,mezzapesa_direct_2012,mezzapesa_line_2013,demir_application_2015,demir_evaluation_2016}. The same technique has been employed also to explore the characteristics of plasma in the plume generated during the ablation process \cite{colombo_self-mixing_2017}. Such possibility of using SMI to gain knowledge about the ablation plume properties is particularly interesting, since it can be essentially useful as an indicator for the process quality. However, for this purpose, further experimental analysis and physical interpretations are required due to the complexity of the plume dynamics, and specific models should be developed to link the interferometric measurement with the process parameters.

In the current paper SMI is proposed as a method for probing the formation and evolution of the plume during multipulse laser ablation, with the aim of monitoring the microdrilling process. A simplified model is presented for describing the optical path difference sensed by a probe laser beam in terms of vapor density and refractive index variations within the expanding plume, linking it to the number of drilling pulses. A self-mixing interferometer based on a diode laser is integrated coaxially with the nanosecond pulsed laser of a microdrilling setup. A digital algorithm for the interferometric signal analysis is developed to demonstrate the possibility of automatic inline monitoring. The experimental results and the morphological analysis of the blind holes obtained for different materials and pulse numbers suggest that the interferometer probes the effective amount of ablated material in the plume. Such a configuration might be used for keeping track of the ablation quality related to the material vaporization rate, in concurrence with the melt-solidification phenomena typical of the thermal interaction given by nanosecond laser pulses.

\section{Model for the SMI signal probing the ablation-induced plume expansion}
\label{sect:path-model}
\begin{figure}[ht]
	\begin{center}
		\includegraphics[width=0.525\linewidth]{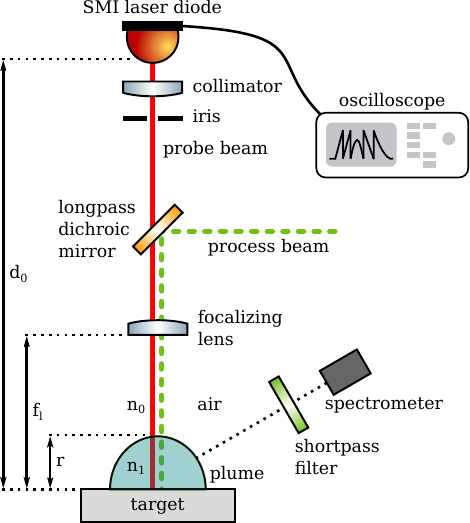}
	\end{center}
	\caption 
	{ \label{fig:setup} Scheme of the main optical parts of the microdrilling and interferometric setups.}
\end{figure}

The measurement principle of a laser interferometer is based on the detection of constructive and destructive interference fringes, obtained by overlapping two or more coherent beams passing through different optical paths. In the classical Michelson interferometer a laser beam is split into two parts, passing through the reference and the sample arms, and then reflected back and recombined on a detector. The self-mixing configuration simplifies such an arrangement, exploiting the optical feedback effect occurring when part of the emitted light gets reflected and interferes within the active medium of the laser source, with fringes being visible as laser intensity modulations \cite{taimre_laser_2015}. A variation in the refractive index or in the physical length of the optical path results in a fringe shift. If $N_f$ is the fringe number, the corresponding optical path difference $\Delta p$ is
\begin{equation}
\Delta p = N_f \frac{\lambda_0}{2}\,,
\end{equation}
where $\lambda_0$ is the wavelength of the laser beam \cite{donati_laser_1995,zabit_self-mixing_2013}. In the current work a self-mixing interferometer is implemented in a microdrilling setup, with a diode laser beam probing the same optical path of the high-power pulsed process laser, as sketched in Fig.~\ref{fig:setup}. In the following, a model for the optical path probed during laser ablation is provided.

\subsection{Single-pulse model}
The interferometer measures an optical path variation, calculated as the difference between the initial and unperturbed optical path length $p_i$, and the optical path $p_f(t)$ in the presence of an ablation plume after a laser pulse \cite{colombo_self-mixing_2017}, defined respectively as
\begin{equation}
p_i = n_0 d_0
\end{equation}
and
\begin{equation}
p_f (t) = n_0\,(d_0-r(t)) + n_1(t)\, r(t)\,.
\end{equation}
Here $n_0$ and $n_1(t)$ are the effective refractive indices of the surrounding gas and of the plume respectively, $d_0$ is the constant distance between the interferometer diode facet and the target plane, i.e., the external cavity length, and $r(t)$ is the extension of the plume region crossed by the probe beam. The time dependence of $n_1(t)$ and $r(t)$ is referred to the start of the ablation process, happening at $t=0$. Therefore the optical path difference $\delta p(t)$ is
\begin{equation}
\label{eq:delta-p-zero}
\delta p(t) = p_f(t)-p_i = r(t)\,(n_1(t)-n_0)\,.
\end{equation}

Since the SMI signal is going to be studied on the millisecond timescale, it can expected that the typical size of the plume is greater than a few millimeters \cite{geohegan_fast_1992,harilal_internal_2003,breitling_fundamental_2004,colombo_self-mixing_2017}, a scale that is much bigger than the ablation crater diameter of the order of \SI{10}{\micro\meter} to \SI{20}{\micro\meter}. Hence, in the first approximation, each ablation pulse can be considered as a point-like instantaneous explosion as in the Sedov--Taylor theory \cite{grun_observation_1991,gupta_direct_1991,callies_time-resolved_1995}. For a simplified model the following assumptions are introduced:
\begin{itemize}
	\item the plume vapor is distributed within the volume delimited by the expanding shock wavefront;
	\item the vapor density in the plume volume is dilute and uniform.
\end{itemize}

For a spherical expansion, the radius $r(t)$ of the shock wavefront predicted by the blast theory \cite{freiwald_approximate_1975} is equal to
\begin{equation}
\label{eq:radius-wave}
r(t) = \frac{\xi_0 E_0^{1/5}}{\rho_0^{1/5}} t^{2/5}\,,
\end{equation}
where $\xi_0$ is an adimensional constant close to one, $\rho_0$ is the density of the unperturbed ambient gas, and $E_0$ is the shock wave energy, which is typically a fraction of the laser pulse energy  \cite{zeng_energy_2004,konig_plasma_2005,amer_shock_2008}. The unperturbed gas can be taken as air at standard temperature and pressure, with $\rho_0\simeq\SI{1.2}{\kilogram\per\meter\cubed}$. Moreover, for low energetic pulses of the order of a few microjoules, \SI{20}{\micro\joule} for the system used for the current measurements, $E_0$ can be estimated from other experimental studies between $0.1\%$ and $1\%$ of the pulse energy \cite{porneala_time-resolved_2009,demir_investigation_2015}.

Consequently, the plume density $\rho_1(t)$, averaged in the region delimited by $r(t)$ and by the target plane, scales inversely with the expanding plume volume $V_1(t)$ as
\begin{equation}
\label{eq:rho-one}
\rho_1(t) = \frac{M_1}{V_1(t)} = \frac{m_1}{r^{3}(t)}\,,
\end{equation}
where $M_1$ is the material mass that gets vaporized and ejected from the target by a single laser pulse, while $m_1$ depends on $M_1$ and on its geometrical distribution in the plume. The angular distribution of the plume is typically concentrated within an angle $2 \theta_p$ along the ablation direction, depending on several factors, such as material type, pulse duration, and ambient gas pressure \cite{amoruso_characterization_1999,verhoff_angular_2012}. Therefore, assuming a homogeneous mass distribution within a conical spherical sector, expanding with a spherical scaling from the ablation crater, the effective mass $m_1$ becomes
\begin{equation}
\label{eq:mass-distribution}
m_1 = \frac{3 M_1}{2\pi(1-\cos{\theta_p})}\,,
\end{equation}
with $\theta_p$ the cone half angle of the plume distribution relative to the normal of the target surface. The latter can be estimated from literature as ranging from an almost hemispherical distribution, with $\theta_p \sim \ang{90}$ and $m_1 = \frac{3}{2\pi} M_1$, to strongly elongated plume jets, with $\theta_p \sim \ang{10}$ \cite{toftmann_angular_2000-1,amoruso_thermalization_2004,colombo_self-mixing_2017}.

It might be assumed that the refractive index $n_1(t)$ probed by the SMI beam in the plume volume can be modeled with the Gladstone--Dale relation for a homogeneous gas \cite{walkup_studies_1986,ventzek_laserbeam_1991,tzortzakis_femtosecond_2001,amer_laser-ablation-induced_2009,sangines_two-color_2011,boudaoud_using_2015}. Therefore, under the hypothesis of a uniform plume density $\rho_1(t)$ within a volume of radius $r(t)$, the index of refraction in the plume region is
\begin{equation}
\label{eq:refractive-index}
n_1(t) = 1+K \rho_1(t) = 1+\frac{K m_1}{r^{3}(t)}\,,
\end{equation}
where $K$ is the Gladstone--Dale constant \cite{gladstone_researches_1863}, which depends on the physical properties of the media. The plume is composed of a mixture of media, and the respective Gladstone--Dale constant $K$ probed by the SMI beam cannot be easily determined neither experimentally nor from literature. However, the value of $K$ for typical combustion neutral gases can be considered as a rough estimation, ranging between \SI{0.2e-3}{\meter\cubed\per\kilogram} and \SI{0.5e-3}{\meter\cubed\per\kilogram} \cite{gardiner_refractivity_1981}.

By inserting \eqref{eq:refractive-index} in \eqref{eq:delta-p-zero} the optical path difference obtained after a single ablation pulse becomes
\begin{equation}
\delta p(t) = r(t)\, \left(1-n_0+\frac{K m_1}{r^{3}(t)}\right) 
=r(t)\, (1-n_0) + \frac{K m_1}{r^{2}(t)}\,.
\end{equation}
If the refractive index of the background gas can be approximated as $n_0\simeq 1$, taking the expression of \eqref{eq:radius-wave} for $r(t)$ the optical path difference scales in time as a power-law equal to
\begin{equation}
\label{eq:delta-p-one}
\delta p(t)\simeq \frac{K  m_1 \rho_0^{2/5}}{\xi_0^2 E_0^{2/5}} t^{-4/5}=\varepsilon t^{-4/5}\,,
\end{equation}
defining
\begin{equation}
\varepsilon=\frac{K m_1 \rho_0^{2/5}}{\xi_0^2 E_0^{2/5}}
\end{equation}
as a quantity that summarizes the characteristics of the ablation induced by a single laser pulse.

\subsection{Multiple pulse model}
The examined microdrilling process consists of a series of $N_p$ consecutive laser pulses, each one of duration $\tau_p$ and with a pulse repetition rate $f_p=t_p^{-1}$. The total process time depends on the number of pulses $N_p$, and is equal to $T_p=N_p t_p$. A preliminary characterization of the laser source used for the following experimental investigation highlighted that the laser pulses are less energetic in the first part of the emission, with a not negligible transition interval to the condition of steady emission \cite{furlan_sub-micrometric_2017}. Therefore, without loss of generality, an offset $N_0$ in the pulse number is introduced, with an effective number of ablation pulses equal to $N_p-N_0$.

For the ablation conditions considered in the current work the laser emission parameters are such that $T_p \gg t_p \gg \tau_p$. Moreover, other experimental studies showed that the dynamics of the ejected plume is turbulent during multipulse ablation, suggesting that the ablated vapor gets mixed by the successive pulses and that the total amount of ablated material within the plume increments gradually during the process \cite{breitling_plasma_2003,breitling_material-vapor_2003,breitling_fundamental_2004,schille_characterisation_2012}. Two further hypotheses are therefore introduced:
\begin{itemize}
	\item the ablation process is approximated as a continuous succession of infinitesimally short and instantaneous pulses;
	\item the vapor ejected after each single pulse accumulates within the plume volume generated by the previous pulses and delimited by $r(t)$.
\end{itemize}

Under the previous assumptions, the overall optical path difference $\Delta p$ probed by the interferometer after $N_p$ pulses can be seen as the linear superimposition of small contributions, each one defined by $\delta p(t)$ of \eqref{eq:delta-p-one} normalized to the single pulse period $t_p$. Therefore $\delta p(t)/t_p$ can be integrated between $t=0$ and $t=T_p$, obtaining
\begin{equation}
\label{eq:deltap-p-integral}
\Delta p \simeq \int^{T_p}_{0}\frac{\delta p(t)}{t_p} \, dt = \frac{\varepsilon}{t_p} \int^{T_p}_{0}t^{-4/5}\, dt = \frac{5\varepsilon}{t_p} T_p^{1/5}\,.
\end{equation}
This expression can be rewritten as a function of the number of pulses as
\begin{equation}
\label{eq:delta-p-two}
\Delta p \simeq \eta \,(N_p-N_0)^{1/5}\,,
\end{equation}
where the amplitude is described by the characteristic length $\eta$, defined as
\begin{equation}
\label{eq:delta-p-eta}
\eta = \frac{5\varepsilon}{t_p^{4/5}} = \frac{5 K m_1 \rho_0^{2/5}}{\xi_0^2 E_0^{2/5}t_p^{4/5}}\,.
\end{equation}
This quantity is essentially determined by the laser parameters, and by the physical properties of the target material and of the plume vapors.

\begin{figure}[ht]
	\begin{center}
		\includegraphics[width=0.7\linewidth]{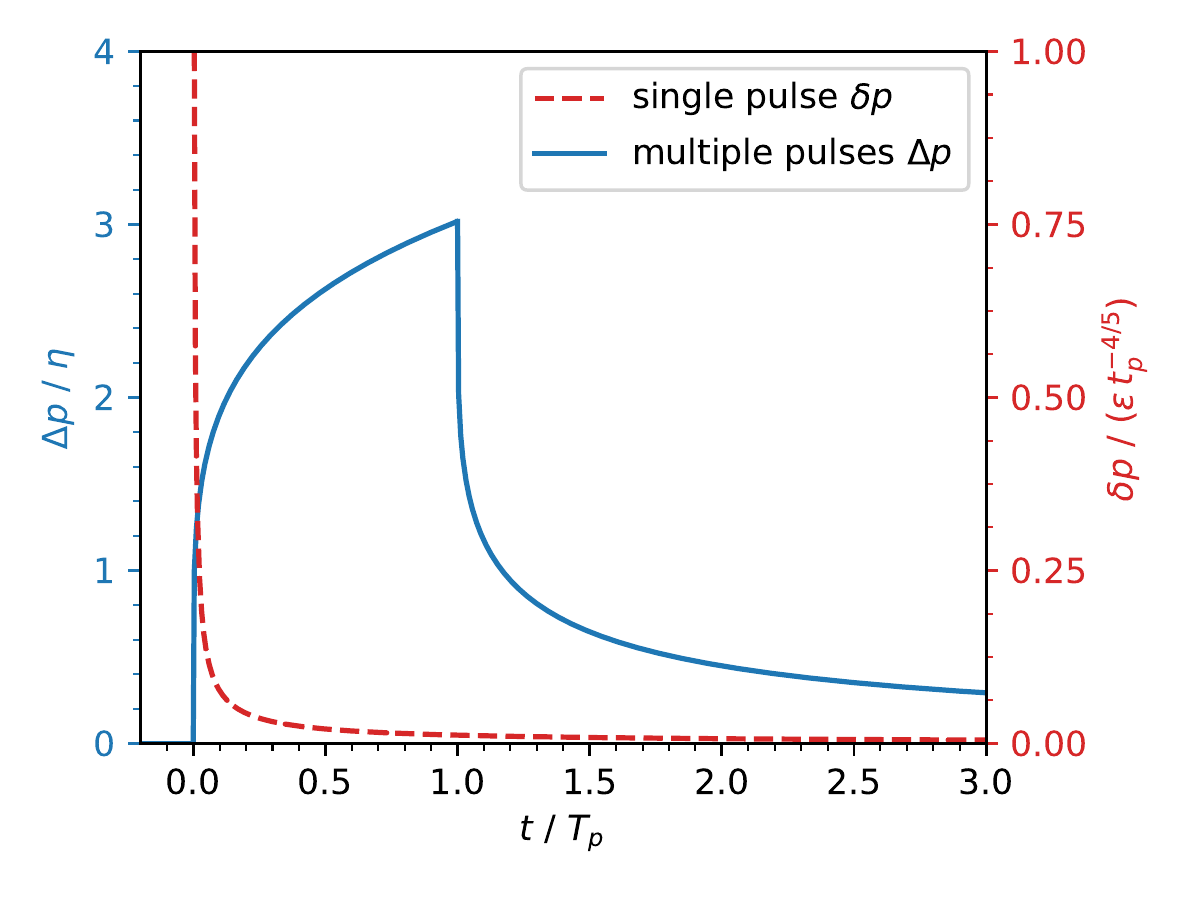}
	\end{center}
	\caption 
	{ \label{fig:simulation} Optical path difference $\Delta p(t)$ calculated with \eqref{eq:delta-p-t} in relative units. The ablation process occurs between $t=0$ and $T_p=N_pt_p$ with $N_p=250$ pulses. The single-pulse contribution $\delta p(t)$ is reported for comparison, calculated with \eqref{eq:delta-p-one}.}
\end{figure}

Writing \eqref{eq:deltap-p-integral} in a general form, the optical path signal $\Delta p(t)$ at time $t$, whose calculated behavior is reported in Fig.~\ref{fig:simulation}, is expected to undergo distinct trends as follows:
\begin{subequations}
\label{eq:delta-p-t}
\begin{align}
\begin{split}
\Delta p(t) = 0 & \qquad t \leq 0
\end{split}\\
\begin{split}
\Delta p(t) = \eta \left(\frac{t}{t_p}\right)^{1/5} & \qquad 0<t \leq T_p
\end{split}\\
\begin{split}
\Delta p(t) = \eta \left(\frac{t}{t_p}\right)^{1/5} -  \eta\left(\frac{t-T_p}{t_p}\right)^{1/5} & \qquad t > T_p\,.
\end{split}
\end{align}
\end{subequations}
Conversely to the single-pulse case, where the optical path $\delta p(t)$ decays in time as calculated in \eqref{eq:delta-p-one}, in the multipulse process the optical path $\Delta p(t)$ increases as a power-law for $t\leq T_p$. This can be explained considering that in such an interval the plume formation is supported by several consecutive pulses, tending to an equilibrium condition as $N_p$ becomes high. After the end of the ablation process, $\Delta p(t)$ decays in time analogously to $\delta p(t)$.

\section{Experimental setup}
\subsection{Microdrilling setup and self-mixing interferometer}
The experimental setup is sketched in Fig.~\ref{fig:setup}, and is the same described in previous works \cite{colombo_self-mixing_2017,donadello_evolution_2018-1}. The processing fiber laser (\textsc{IPG Photonics YLPG-5}) emits at \SI{532}{\nano\meter} with an average power of \SI{6}{\watt}. The process beam is deflected by \ang{90} toward the target specimen with a longpass dichroic mirror (\textsc{Thorlabs DMLP567}) having \SI{567}{\nano\meter} cutoff. The laser parameters are reported in Table~\ref{tab:parameters}. The pulse duration is $\tau_p=\SI{1.2}{\nano\second}$ with a peak power of \SI{16}{\kilo\watt}. The pulse frequency is fixed to $f_p=t_p^{-1}=\SI{160}{\kilo\hertz}$, while the number of drilling pulses $N_p$ is controlled by varying the emission duration $T_p=N_p t_p$ between \SI{0.3}{\milli\second} ($N_p=50$) and \SI{1.6}{\milli\second} ($N_p=250$).

\begin{table}[ht]
	\caption{Characteristics of the pulsed microdrilling laser and of the continuous probe laser.}
	\label{tab:parameters}
	\begin{center}
		\begin{tabular}{lll}
			\hline
			Parameter & Process laser & SMI laser \\
			\hline
			wavelength & \SI{532}{\nano\meter}  & \SI{785}{\nano\meter}  \\
			power & \SI{16}{\kilo\watt} {(peak)} & \SI{15}{\milli\watt} \\
			pulse energy & \SI{20}{\micro\joule} & -- \\
			pulse duration $\tau_p$ & \SI{1.2}{\nano\second} & -- \\
			pulse repetition rate $f_p$ & \SI{160}{\kilo\hertz} & -- \\
			pulse number $N_p$ & $50$ -- $250$ & -- \\
			lens focal length $f_l$ & \SI{100}{\milli\meter} & \SI{100}{\milli\meter}  \\ 
			focused beam diameter & \SI{22}{\micro\meter} & \SI{41}{\micro\meter}/\SI{24}{\micro\meter} \\ \hline
		\end{tabular}
	\end{center}
\end{table}

The self-mixing interferometer is based on a GaAlAs multi-quantum well laser diode with built-in monitor photodiode (\textsc{Hitachi HL7851G}), emitting \SI{15}{\milli\watt} at $\lambda_0 = \SI{785}{\nano\meter}$ and collimated with a \SI{10}{mm} lens. The diode is placed at distance $d_0=\SI{410}{\milli\meter}$ from the target specimen. The choice of the diode wavelength is convenient since the plasma emission in the spectral interval around $\lambda_0$ is typically weak \cite{demir_application_2015}. In fact, a strong emission would increase the possibility of laser mode hopping and intensity modulations, even if the process radiation emitted at the laser wavelength would add incoherently to the laser field without altering the interferometric fringe signal.

The wavelength of the probe beam is transmitted by the dichroic mirror, allowing it to be superimposed on the process beam. The two beams are then focalized on the target surface by means of an achromatic doublet lens (\textsc{Thorlabs AC254-100-A-ML}) with focal length $f_l=\SI{100}{\milli\meter}$. In the focus, the calculated process beam diameter is \SI{22}{\micro\meter}, while the expected fast and slow axes of the elliptical interferometer beam are \SI{41}{\micro\meter} and \SI{24}{\micro\meter} respectively. Experiments were carried out in ambient atmosphere without the use of assist gas. Previous investigations showed that, in the absence of any gas flow blowing the plume from the side, in the considered conditions the SMI beam interacts mainly with the plume rather than measuring the displacement of the ablation front \cite{colombo_self-mixing_2017}.

The fraction of the probe light that gets scattered or reflected back to the laser cavity is the origin of the self-mixing phenomenon, introducing a modulation in the laser field intensity and frequency \cite{giuliani_laser_2002}. Indeed, variations in the optical path cause the appearance of interferometric fringes in the signal detected by the monitor photodiode of the laser, with the signal behavior being determined by the optical feedback parameter labeled with $C$ \cite{yu_optical_2009}. The amount of light coupled back to the laser cavity, and consequently $C$, can be limited by regulating the clear aperture of an iris crossed by the collimated SMI beam. Accordingly, the interferometer is operated in the moderate coupling regime, thus with $1<C<4.6$, where the interferometric signal is characterized by a sawtooth-like modulation. With higher values of $C$ the signals would exhibit an increased hysteresis, and the correspondent strong feedback regime is avoided since it can lead to laser mode hopping and fringe count losses. The photodiode signal is conditioned by a single-stage transimpedance operational amplifier having \SI{1}{\mega\hertz} bandwidth. About \SI{14}{\milli\second} of the SMI signal during and after each microdrilling run are acquired with a digital oscilloscope (\textsc{Rigol MSO4024}), with \SI{350}{\mega\hertz} bandwidth and \SI{50}{Msps} sampling rate. The recorded series are transmitted through the \textsc{LXI} interface directly to a computer for the signal processing.

Spectroscopy is used to characterize the process optical emission, providing evidence to the eventual formation of plasma. The optical radiation emitted during the ablation process in the spectral range between \SI{300}{\nano\meter} and \SI{500}{\nano\meter} is acquired with a fiber optic spectroscope (\textsc{Avantes AvaSpec-2048}), having a FWHM resolution of \SI{0.8}{\nano\meter} and an integration time set to \SI{2}{\milli\second}. A shortpass optical filter with \SI{500}{\nano\meter} cutoff (\textsc{Edmund 47-287}) suppresses the process laser radiation, and the acquired spectra are corrected according to the filter transmission curve.

\begin{figure}[ht]
	\centering
		\includegraphics[width=0.7\linewidth]{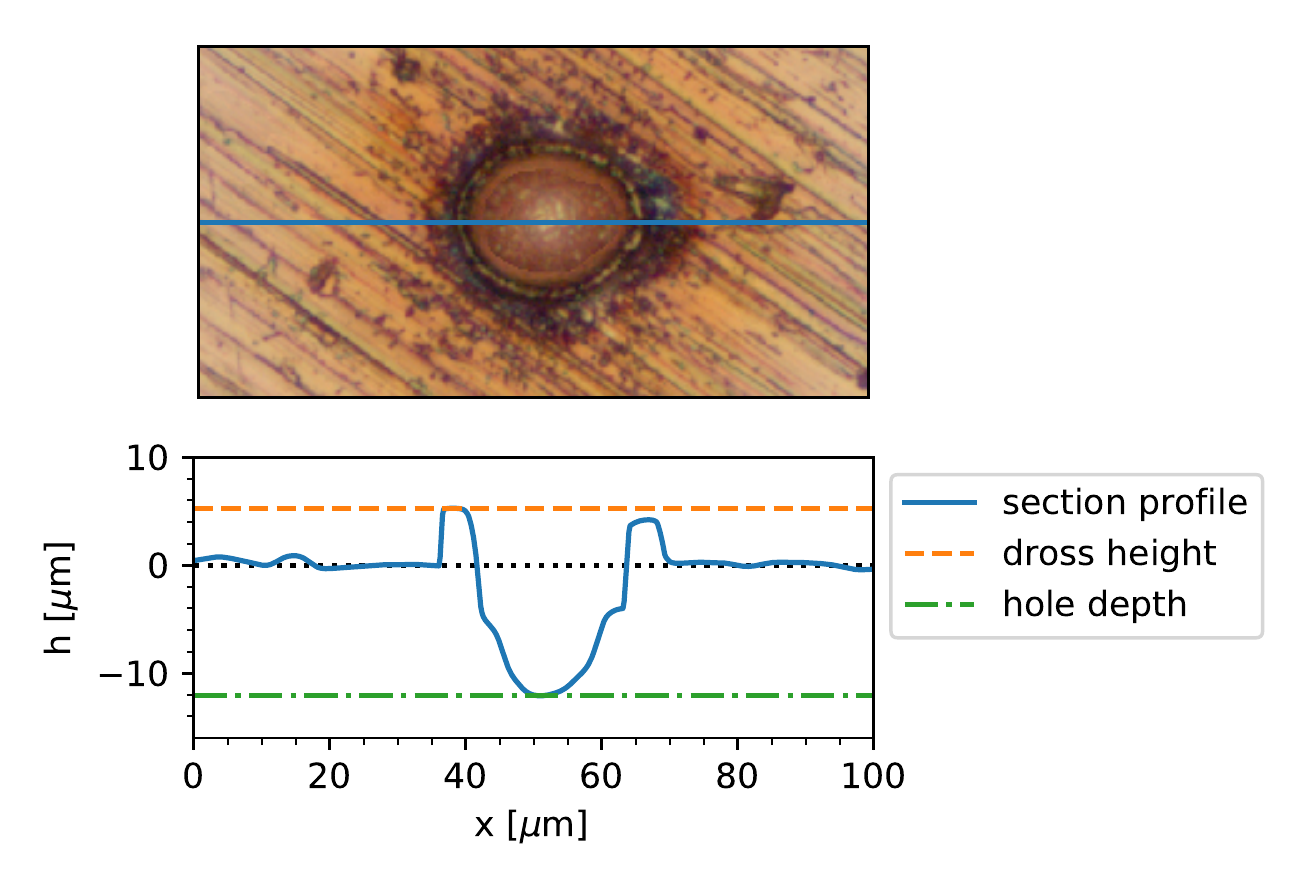}
	\caption 
	{ \label{fig:hole-depth-example} Example of hole drilled on a copper target with $N_p=150$ pulses, acquired with focus variation microscopy. The maximum hole depth and dross height can be measured from the hole profile acquired along a section and reported in the bottom plot.}
\end{figure}

The morphology of the microdrilled blind holes is analyzed by means of a three-dimensional (3D) focus variation microscope (\textsc{Alicona Infinite Focus}). The 3D surface profiles are acquired at $50 \times$ magnification, with vertical and lateral resolutions equal to \SI{50}{\nano\meter} and \SI{1.5}{\micro\meter} respectively. The maximum hole depth $h_\text{hole}$ and dross height $h_\text{dross}$ are measured for each sample relatively to the average surface plane, as in the example reported in Fig.~\ref{fig:hole-depth-example}.

The behavior of the SMI signal is evaluated while microdrilling the following materials:
\begin{description}
	\item[SST (AISI 301)] stainless steel (EN 1.4310),  \SI{0.2}{\milli\meter} thick foil (\textsc{Lamina S.p.A.});
	\item[Cu (110)] copper alloy (ASTM B152), \SI{1}{\milli\meter} thick foil (\textsc{Metal Center S.r.l.});
	\item[Ti] commercially pure grade 2 titanium, \SI{0.3}{\milli\meter} thick foil (\textsc{Lamina S.p.A.});
	\item[TiN] titanium nitride ceramic coating, \SI{3.87}{\micro\meter} thick (\textsc{Lafer S.p.A.}), produced by means of physical vapor deposition (PVD) on steel substrate (39NiCrMo3).
\end{description}
The main physical characteristics of these materials are summarized in Table \ref{tab:materials}.

\begin{table}[ht]
	\caption{Typical characteristics of the materials used as targets for the microdrilling process: density at room temperature $\rho_m$, melting temperature $T_m$, specific heat $c_p$, thermal conductivity $k$, and thermal diffusivity $\alpha$ \cite{pierson_handbook_1996,aws_brazing_2007}.}
	\label{tab:materials}
	\begin{center}       
		\begin{tabular}{llllll}
			\hline
			Material	 & $\rho_m$ & $T_m$ & $c_p$ & $k$ & $\alpha$ \\
			&$\!\left[\frac{\si{g}}{\si{\centi\meter^3}}\right]$&$[\si{\kelvin}]$&$\!\left[\frac{\si{\joule}}{\si{\kilogram\kelvin}}\right]$&$\!\left[\frac{\si{\watt}}{\si{\meter\kelvin}}\right]$&$\!\left[\frac{\si{\milli\meter^2}}{\si{\second}}\right]$\\
			\hline
			SST & $8.03$ & $1693$ & $500$ & $16$ & $4.0$ \\
			Cu & $8.89$ & $1360$ & $385$ & $388$ & $113$ \\
			Ti & $4.51$ & $1964$ & $582$ & $16$ & $6.1$ \\
			TiN & $5.40$ & $2950$ & $545$ & $19$ & $6.5$ \\
			\hline
		\end{tabular}
	\end{center}
\end{table}

\subsection{SMI signal analysis}
\label{sect:signal-analysis}
\begin{figure}[ht]
	\begin{center}
		\includegraphics[width=0.8\linewidth]{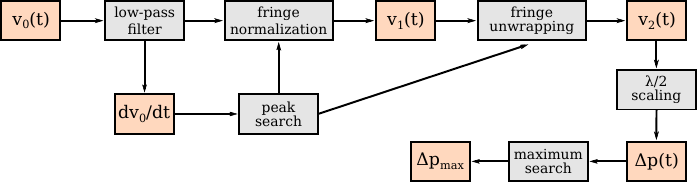}
	\end{center}
	\caption 
	{ \label{fig:algorithm-scheme} Scheme of the algorithm for the interferometric signal analysis.} 
\end{figure}

The SMI voltage signal $v_0(t)$ can contain a variable number of interference fringes, observed as sawtooth-like modulations since the interferometer is operated in the moderate feedback regime. The derivative sign at the signal discontinuity allows the sign of the optical path variation sign, which is origin of the fringe, to be distinguished. To extract the optical path difference $\Delta p (t)$ and its maximum value $\Delta p_{\text{max}}$ from a high number of experimental runs, the signal is analyzed with a digital algorithm written in \textsc{Python}, similar to other fringe unwrapping procedures \cite{norgia_fully_2011,fan_improving_2011,arriaga_real_2012,zabit_self-mixing_2013}. The algorithm takes into account also noninteger fringe numbers, thus improving the measurement resolution below the $\lambda_0/2$ limit, following the steps described below and sketched in Fig.~\ref{fig:algorithm-scheme}:
\begin{enumerate}
	\item A Butterworth low-pass filter with \SI{300}{\kilo\hertz} cutoff suppresses the high-frequency noise.
	\item The discontinuities in the voltage signal $v_0(t)$ are identified by means of a peak search algorithm applied to the time derivative $dv_0/dt$, based on local maxima detection.
	\item The SMI feedback parameter $C$ may change in time during the ablation process due to the mutable optical conditions, such as variations in reflectivity or absorption from the probed system. Therefore $v_0(t)$ needs to be opportunely normalized and translated to a signal $v_{1}(t)$ expressed in fringe number. The signal intervals delimited by two consecutive fringes are normalized to $1$, while the remaining intervals are normalized to the height of nearest fringe discontinuity; if no fringe is present in $v_0(t)$, the whole series is normalized to the average scaling factor of the respective data set.
	\item The unwrapping procedure reconstructs a normalized and continuous signal $v_{2}(t)$ by subtracting the cumulative offset introduced by the fringe discontinuities.
	\item The unwrapped signal $v_{2}(t)$ is converted to optical path difference by knowing that each interference fringe corresponds to a path variation equal to half laser wavelength, i.e., $\Delta p(t) = v_{2}(t)\, \frac{\lambda_0}{2}\,.$
	\item The maximum value of the optical path $\Delta p_{\text{max}}$ is extracted from each series.
\end{enumerate}

\begin{figure}[ht]
\begin{center}
	\includegraphics[width=0.45\columnwidth]{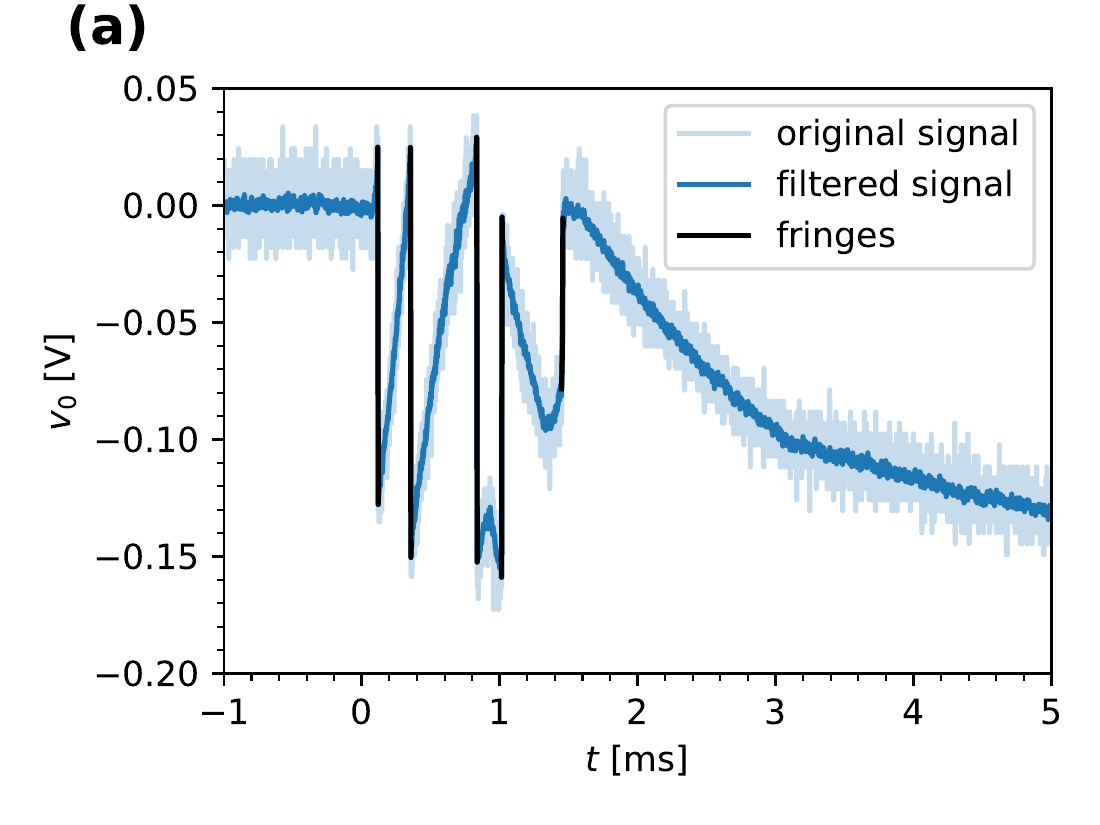}
	\includegraphics[width=0.45\columnwidth]{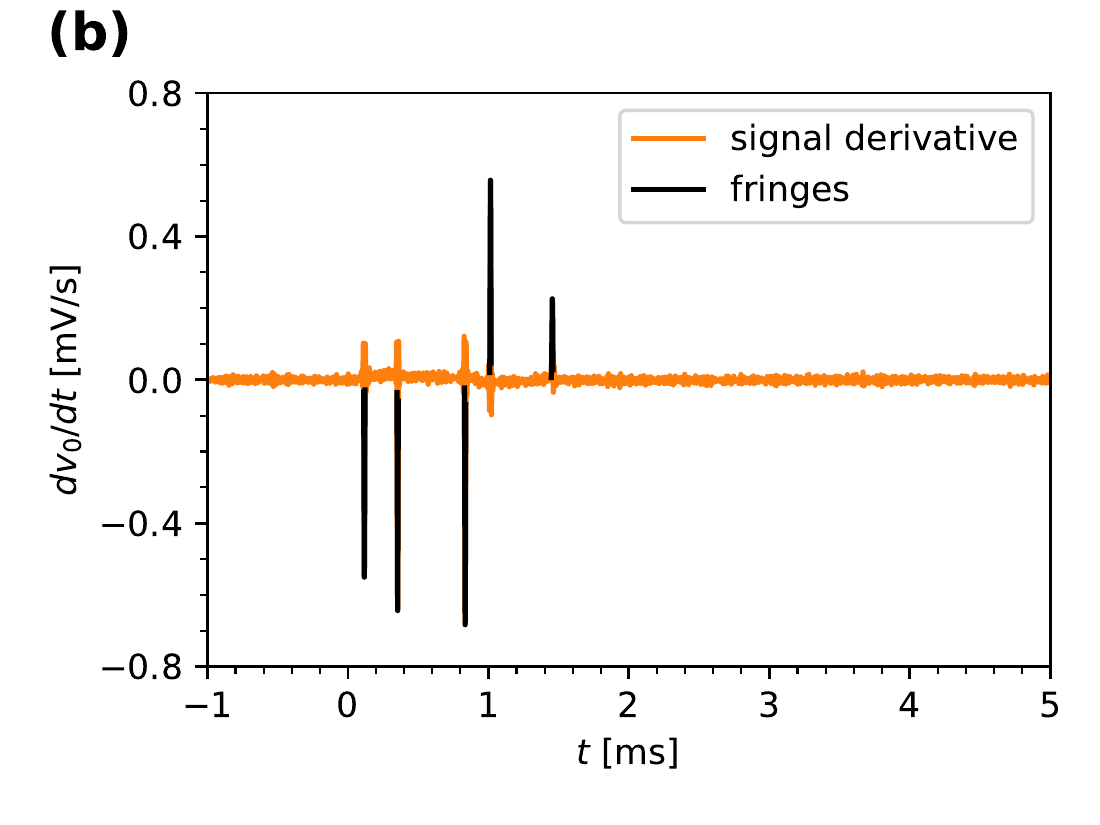}\\[0.5em]
	\includegraphics[width=0.45\columnwidth]{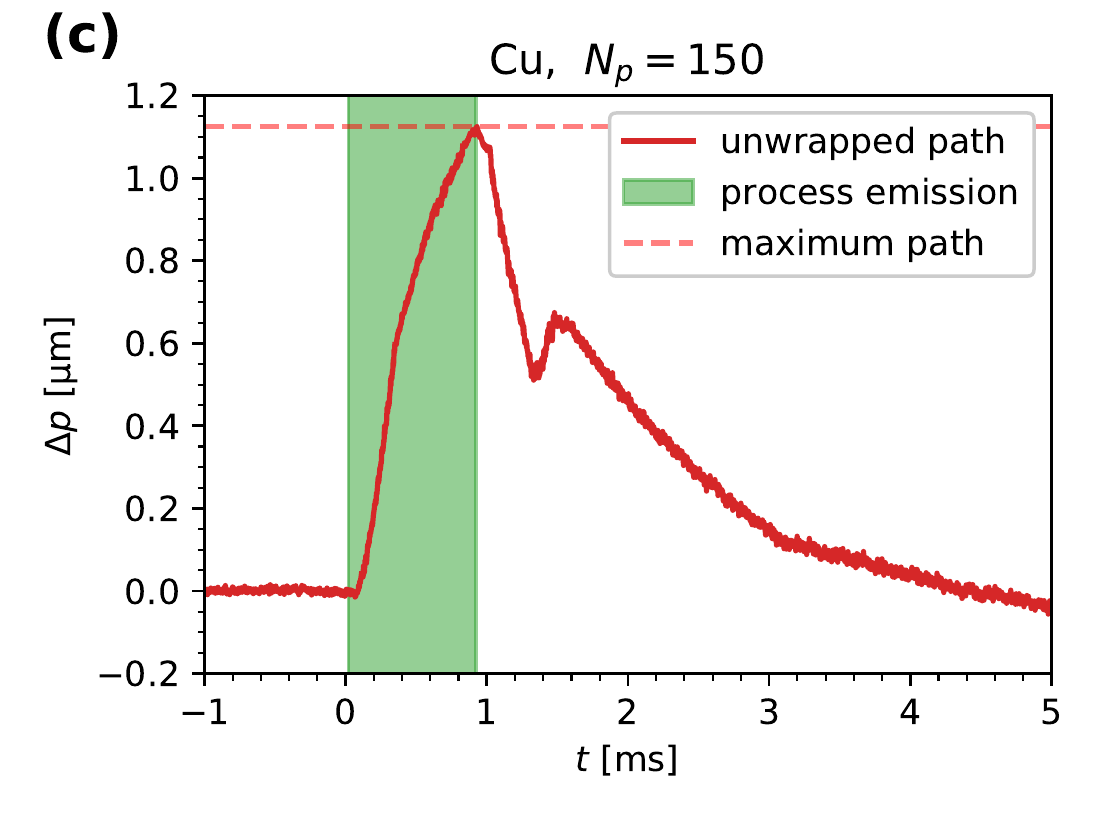}
\end{center}
\caption 
{ \label{fig:signal-example} Plots representing the main steps of the SMI signal analysis: (a) filtering of the high-frequency noise in the voltage signal $v_0(t)$; (b) fringe identification as peaks in the signal derivative; (c) signal normalization, unwrapping, and conversion to optical path difference $\Delta p(t)$. The ablation process starts at $t=\SI{0}{\milli\second}$. The considered material is copper, with $N_p=150$ drilling pulses.}
\end{figure} 

An example of analyzed signal is reported in Fig.~\ref{fig:signal-example}. Several factors may lead to wrong measurement results, such as errors in the fringe detection when the interferometer exits the moderate feedback regime. E.g., this may happen when $C<1$, thus approaching the weak feedback condition with signals tending to asymmetric sinusoidal modulations instead of a sawtooth-like fringes, or when $C>4.6$, where the strong regime can lead to fringe losses \cite{giuliani_laser_2002,yu_optical_2009}. Accordingly, the algorithm identifies and excludes the series that might be misleading for the subsequent analysis, i.e., matching at least one of the following criteria:
\begin{itemize}
	\item the unwrapped signal does not return close to its initial value within a few milliseconds, suggesting a probable failure in the signal acquisition or fringe detection;
	\item signal saturation with strong voltage jumps are observed, probable signature of laser mode hopping when the diode enters the strong feedback regime;
	\item more than a reasonable maximum fringe number is found or the signal-to-noise ratio for $\Delta p_{\text{max}}$ is too low, as it happens in noisy signals.
\end{itemize}
It must be noted that some signals exhibit a cusp-like behavior in the correspondence of fringe discontinuities. This can be an effect of the AC coupling of the photodiode circuit, and it does not have any physical meaning from the interferometric point of view.

\section{Experimental results}
\subsection{Optical path difference}
\label{sect:path}
\begin{figure}[ht]
	\begin{center}
		\includegraphics[width=0.45\linewidth]{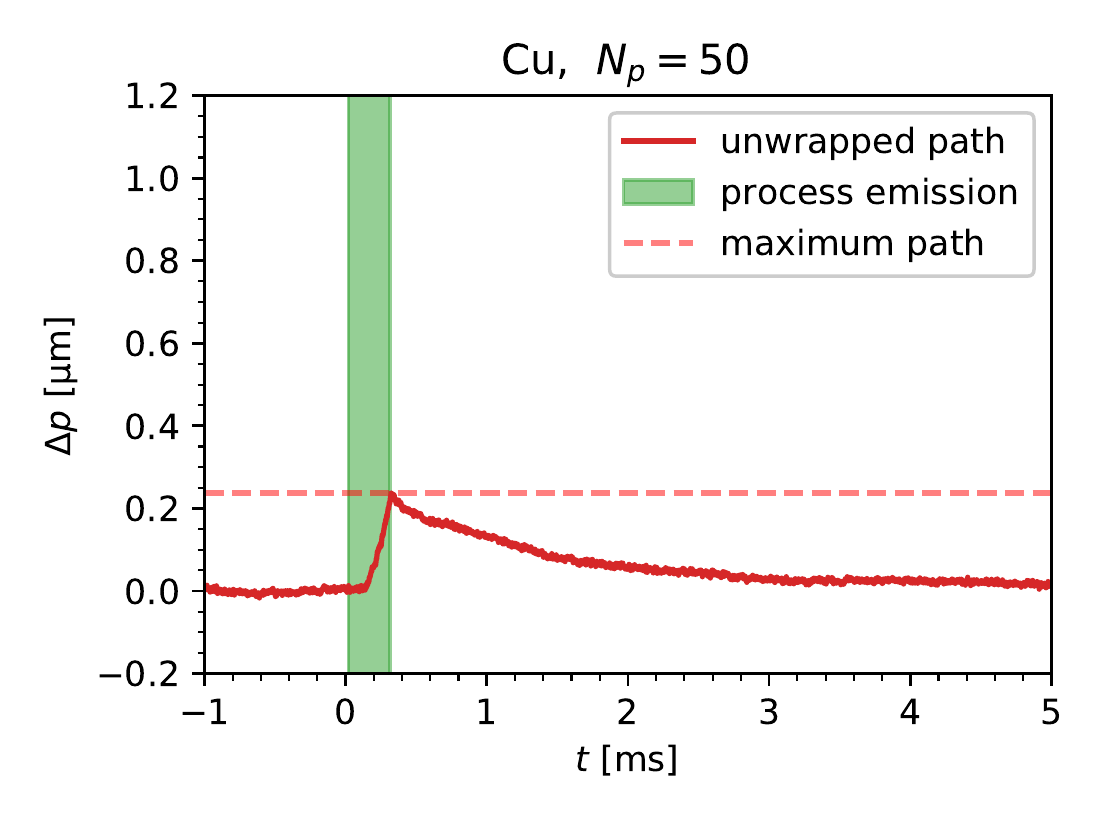}	\includegraphics[width=0.45\linewidth]{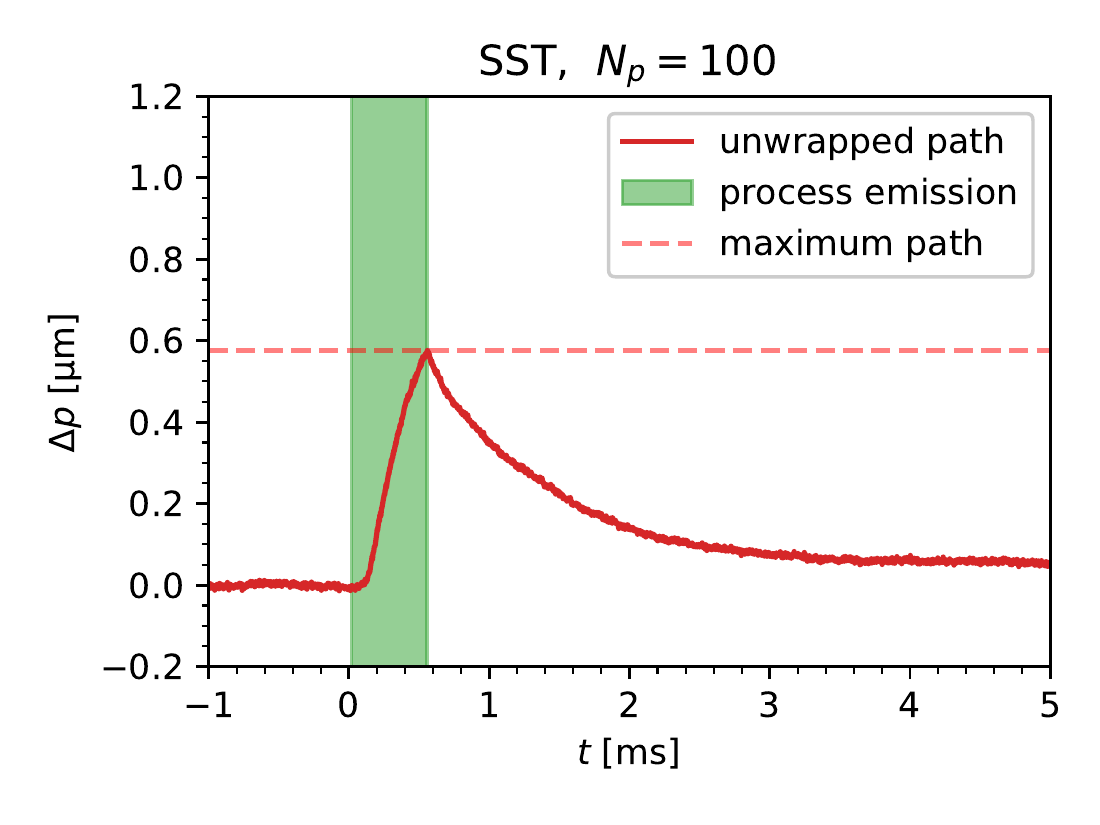}\\
		\includegraphics[width=0.45\linewidth]{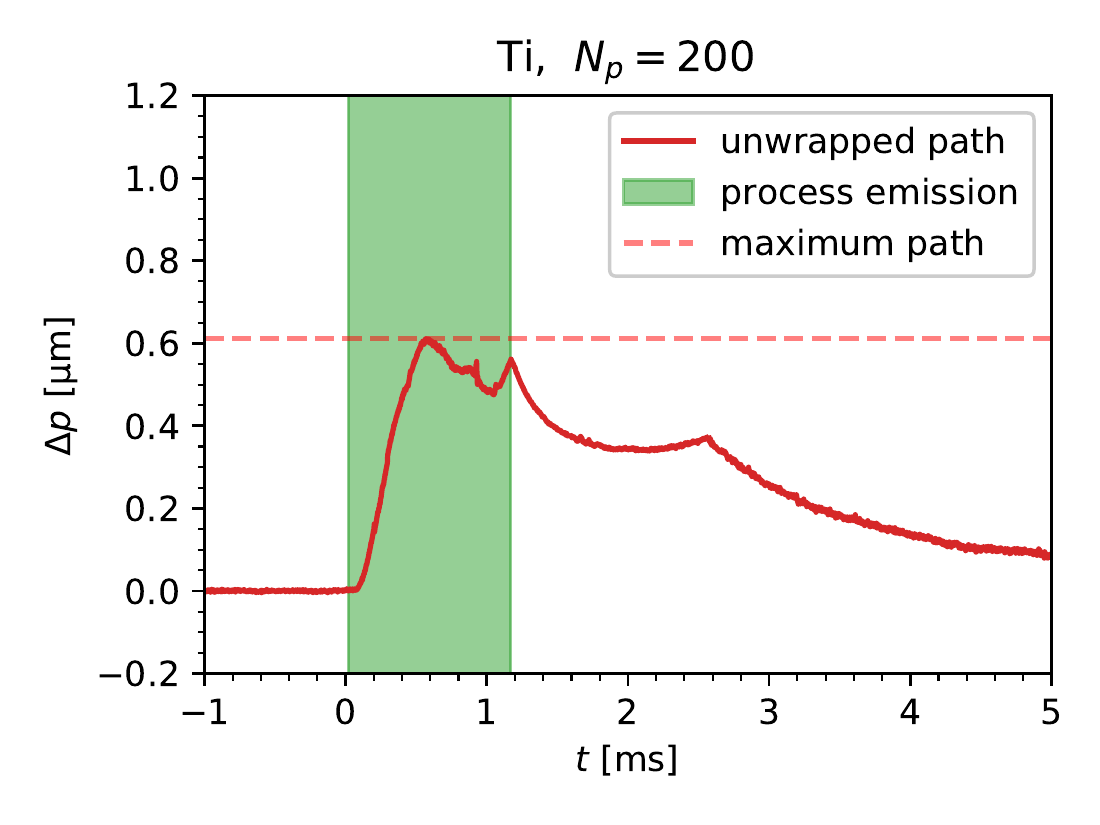}	\includegraphics[width=0.45\linewidth]{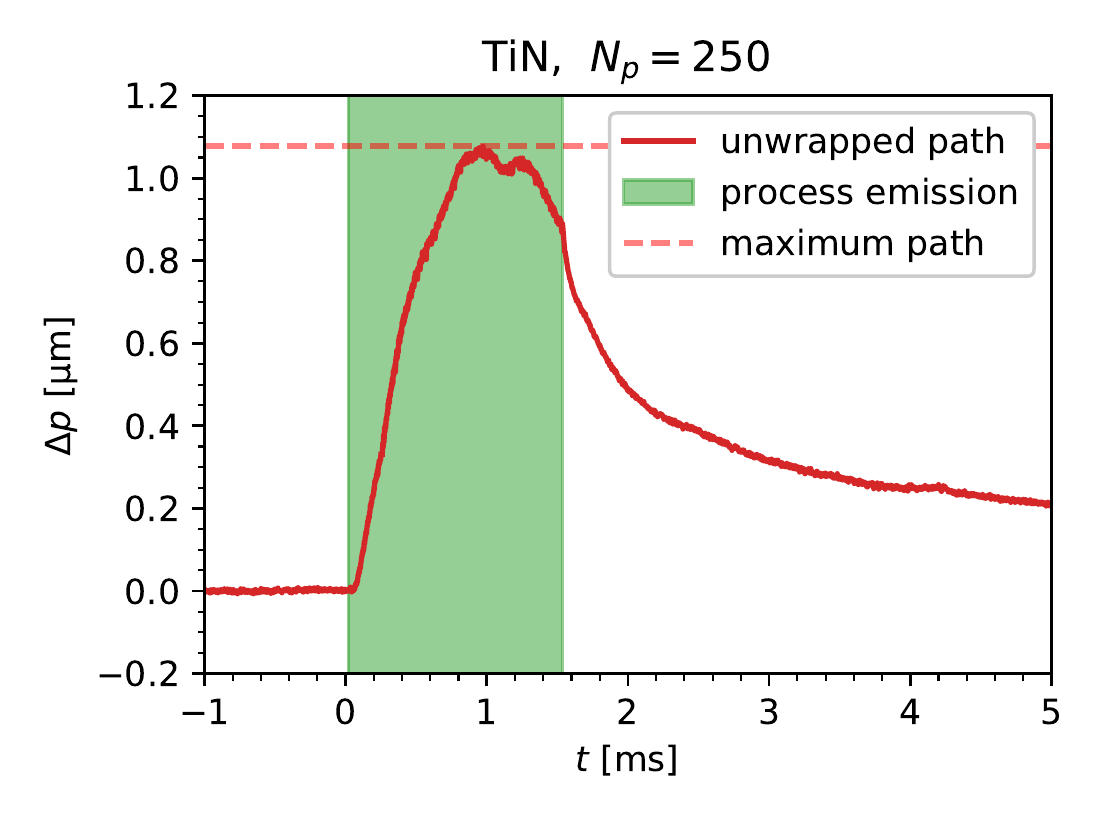}
	\end{center}
	\caption 
	{ \label{fig:path-examples} Optical path difference measured for different target materials and pulse number values, as reported above each plot.}
\end{figure}

The measurements confirmed that the SMI beam interacts with the process plume rather than measuring the drilling depth. In fact, for a typical sequence, the measured optical path difference $\Delta p(t)$ is positive, growing abruptly with the beginning of the microdrilling process on the micrometer scale. For a monotonic signal, such as expected from the model of \eqref{eq:delta-p-t} for $\Delta p(t)$ with $t\leq T_p$, its maximum value after $N_p$ pulses should correspond to $\Delta p (t=t_pN_p)$. However, especially when the number of pulses $N_p$ is high, i.e., the process interval $T_p$ is long, a kind of saturation effect in $\Delta p(t)$ is often observed, reaching its maximum $\Delta p_\text{max}$ at the end of the ablation process or slightly before. After the laser emission has stopped $\Delta p(t)$ decays, returning close to its initial value within few milliseconds. Some examples of signal series are reported in Figures \ref{fig:signal-example} and \ref{fig:path-examples}.

\begin{figure}[ht]
	\begin{center}
			\includegraphics[width=0.8\linewidth]{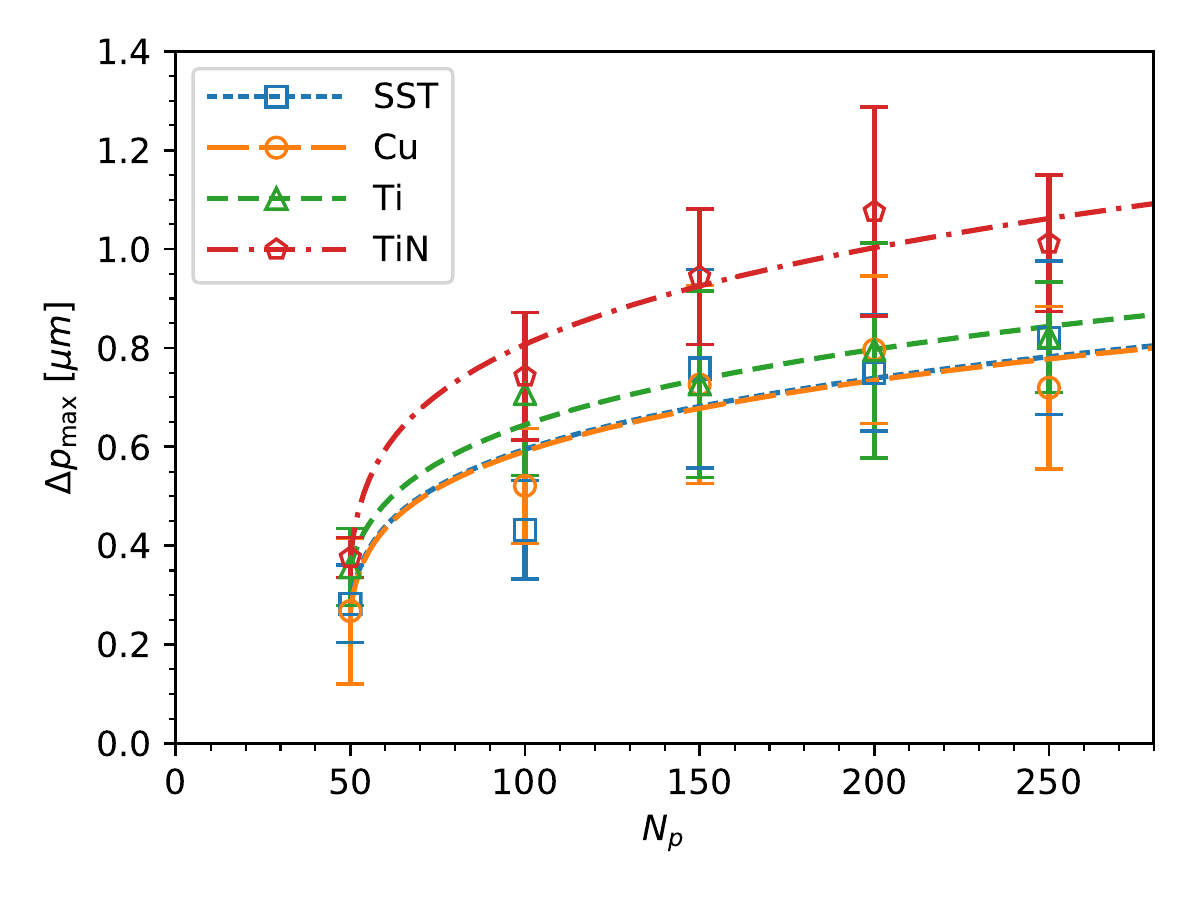}
	\end{center}
	\caption 
	{ \label{fig:max-optical-path-fit} Average maximum value of the optical path difference $\Delta p_{\text{max}}$ as a function of the drilling pulse number $N_p$ for the different materials. The data sets are fitted with \eqref{eq:delta-p-two} (dashed lines).} 
\end{figure}

The maximum of the optical path difference $\Delta p_{\text{max}}$ is measured in the different experimental conditions, with $20$ repetitions for each combination of material and pulse number. About the $32\%$ of the experimental runs are excluded by the signal analysis algorithm, being identified as invalid according to the criteria described before. The average trend of $\Delta p_{\text{max}}$ as a function of $N_p$ is reported in Fig.~\ref{fig:max-optical-path-fit}. The data are fitted with the power-law of \eqref{eq:delta-p-two}. The fitting coefficients $\eta$ and $N_0$ are reported in Table~\ref{tab:path-fit-coeff}. It can be observed that, although stainless steel and copper produce signals with similar amplitudes, for pure titanium $\eta$ is $8\%$ higher, while for titanium nitride it is significantly different, i.e., $36\%$ higher.
 
\begin{table}[ht]
	\caption{Coefficients $\eta$ and $N_0$ of \eqref{eq:delta-p-two} fitting $\Delta p_{\text{max}}$ for the data sets of Fig.~\ref{fig:max-optical-path-fit}, ablation rate $M_1$, expressed as average mass per pulse and calculated from the hole depth measurements of Fig.~\ref{fig:hole-depth}. The values are expressed with their asymptotic standard errors. The respective coefficients of determination $R^2$ of the fitting procedure are also reported.}
	\label{tab:path-fit-coeff}
	\begin{center}       
		\begin{tabular}{lllll}
			\hline
			Material & $\eta$  [\si{\micro\meter}] & $N_0$ &  $R^2$ & $M_1$ [\si{ng}] \\
			\hline
			SST & \num{0.271+-0.009} & \num{48.8+-0.8}  & $0.65$ & \num{0.30+-0.03} \\
			Cu & \num{0.269+-0.008}  & \num{49.0+-0.7}  & $0.58$ & \num{0.27+-0.02} \\
			Ti &  \num{0.292+-0.009} & \num{47.2+-1.6}  & $0.58$ & \num{0.25+-0.01}\\
			TiN & \num{0.368+-0.009} & \num{48.9+-0.7}  & $0.73$ & \num{0.08+-0.02} \\
			\hline
		\end{tabular}
	\end{center}
\end{table}

\subsection{Spectroscopic analysis}
\label{sect:spectra}
\begin{figure}[ht]
	\begin{center}
		\includegraphics[width=0.45\linewidth]{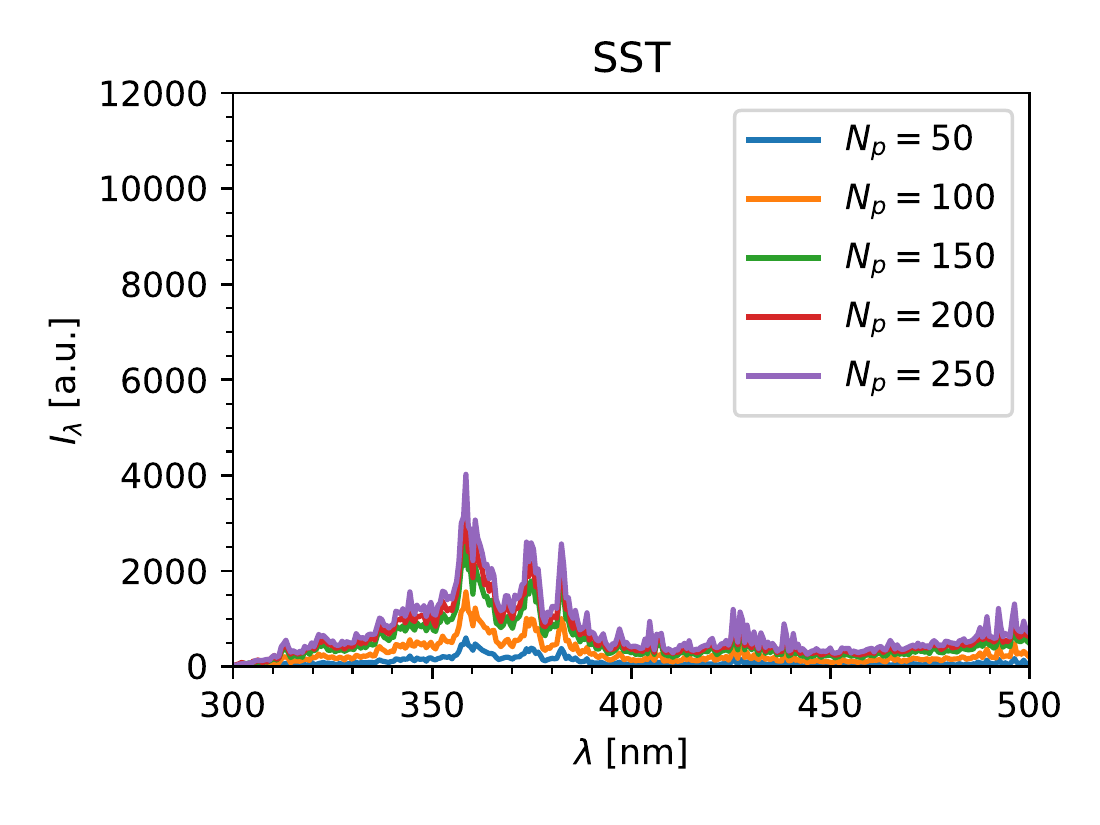}	\includegraphics[width=0.45\linewidth]{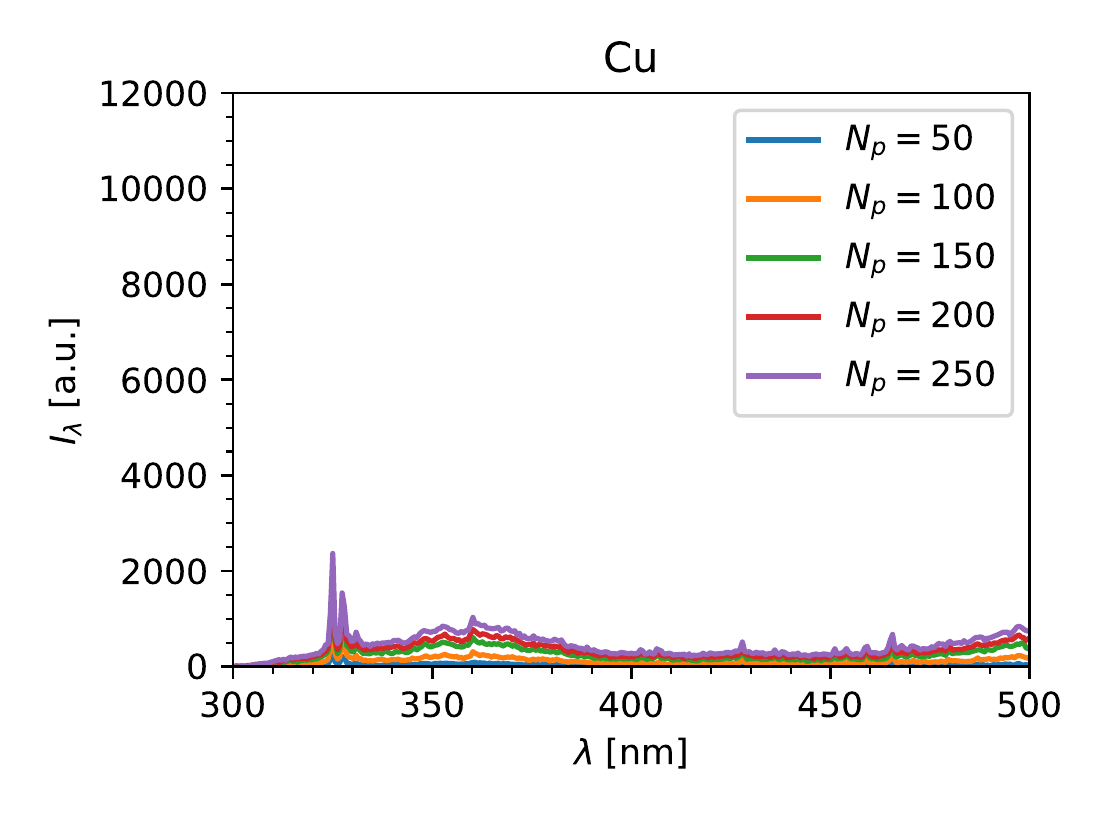}\\	\includegraphics[width=0.45\linewidth]{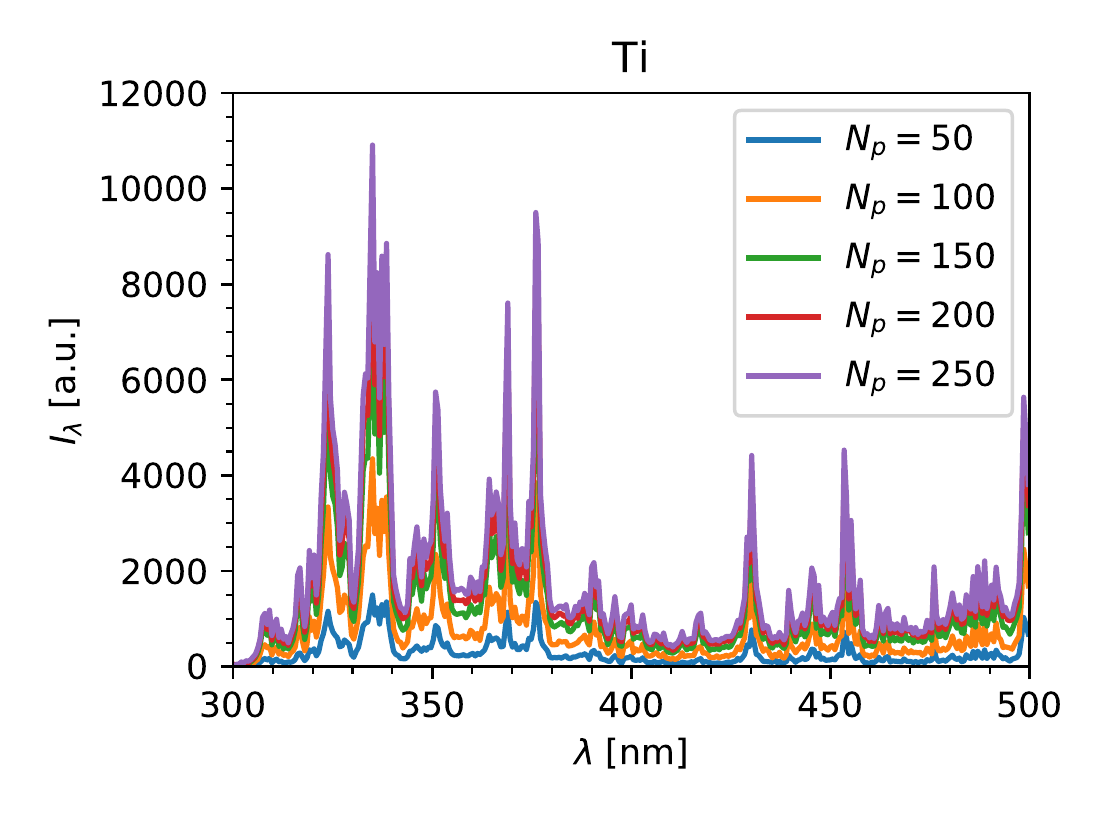}	\includegraphics[width=0.45\linewidth]{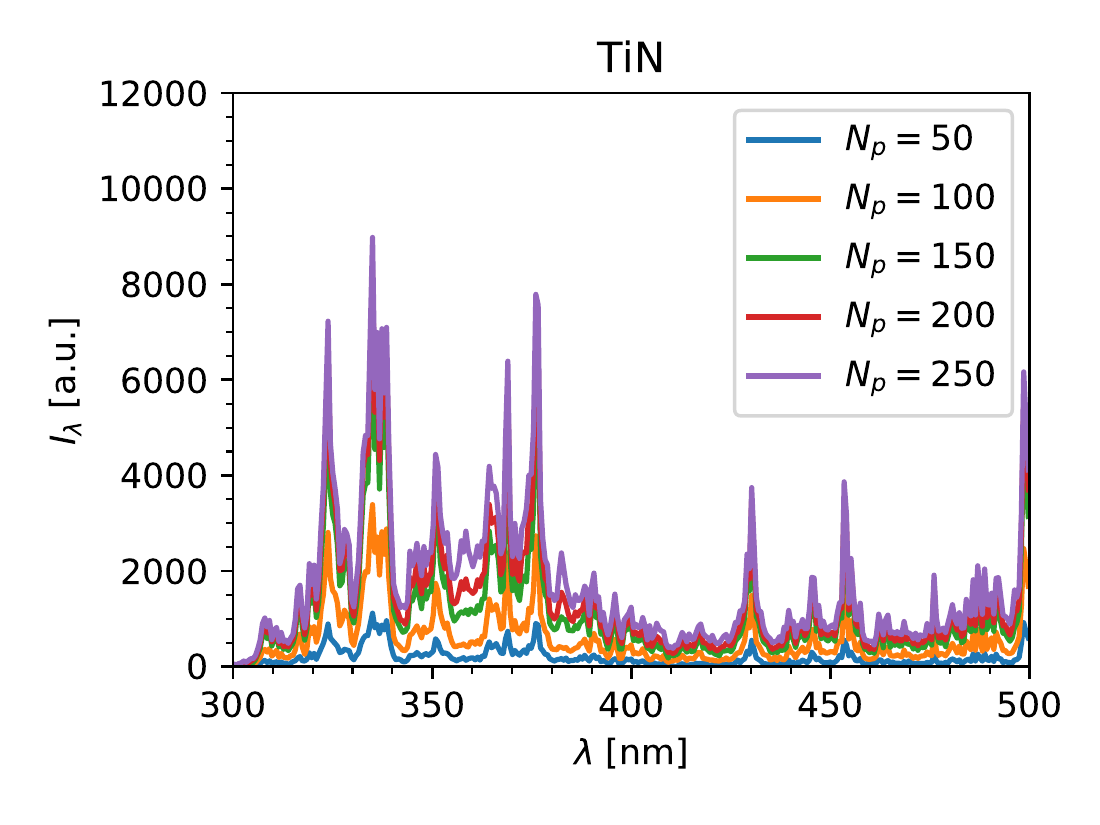}
	\end{center}
	\caption 
	{ \label{fig:spectra-materials} Average spectra of the ablation process emission, grouped by material and pulse number.} 
\end{figure}

The visible emission spectra acquired during the microdrilling process have been grouped and averaged by material and pulse number. The results are reported in Fig.~\ref{fig:spectra-materials}. From a qualitative point of view, the intensity of the spectroscopic peaks of the plasma emission grows with $N_p$ for each material, as expected from an increasing plasma amount. Moreover, different distinctive spectrum shapes can be observed. In particular, Ti and TiN are characterized by several strong discrete lines, while for copper only a few plasma lines are visible, with their intensity comparable to the almost continuous background radiation.

\subsection{Hole morphology analysis}
\label{sect:morphology}
\begin{figure}[ht]
	\begin{center}
		\includegraphics[width=0.8\linewidth]{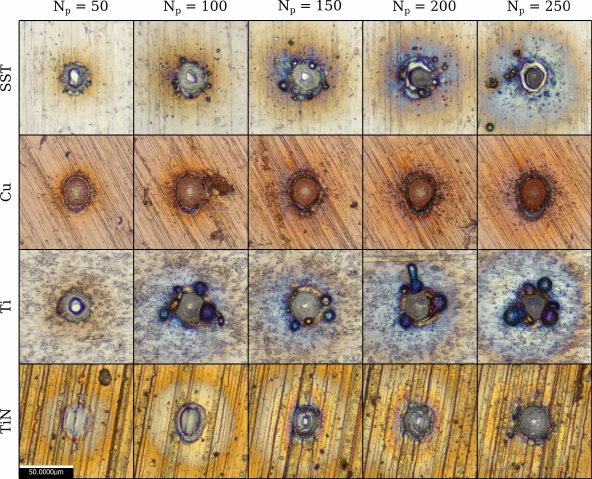}
	\end{center}
	\caption 
	{ \label{fig:hole-images} Examples of microscopy images of microdrilled blind holes, for the different combinations of material and pulse number.} 
\end{figure}

\begin{figure}[ht]
	\begin{center}
		\includegraphics[width=0.45\linewidth]{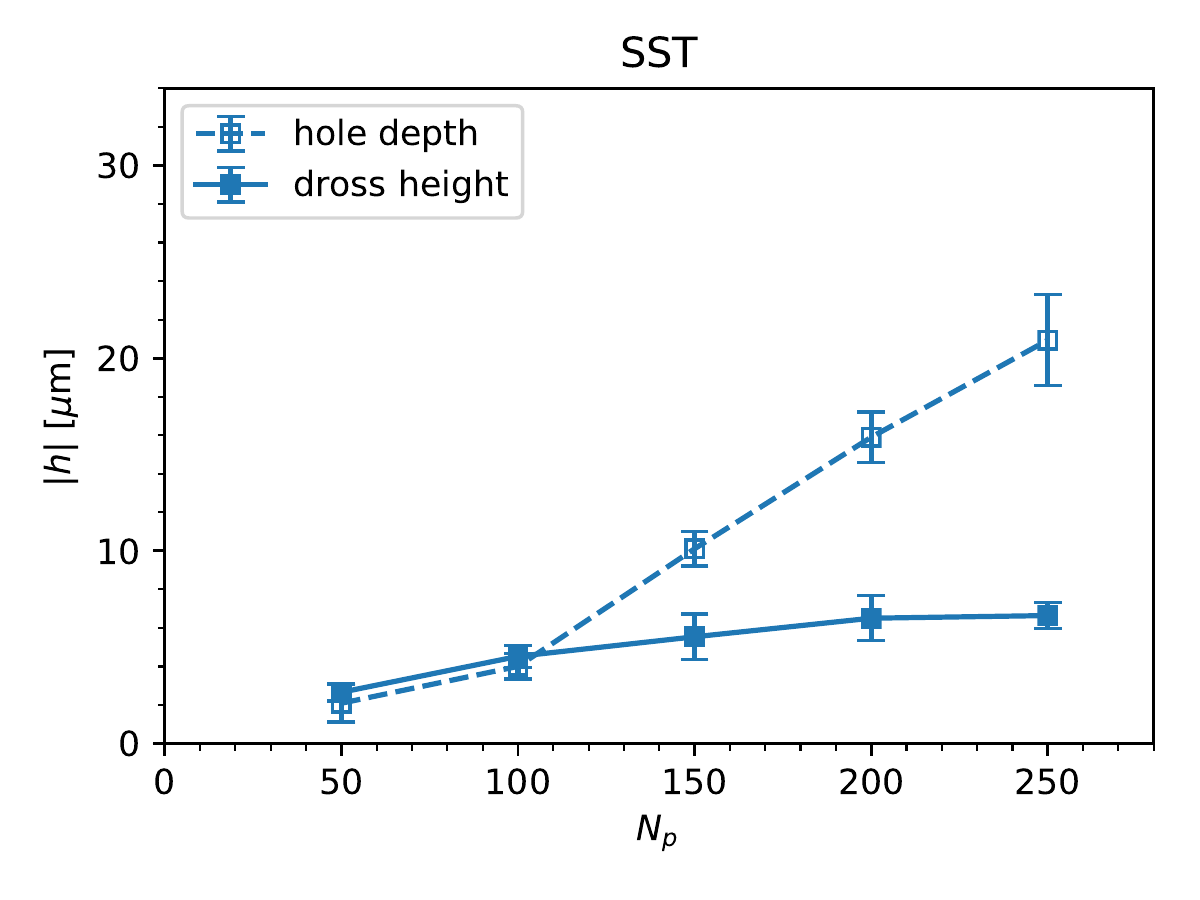}
		\includegraphics[width=0.45\linewidth]{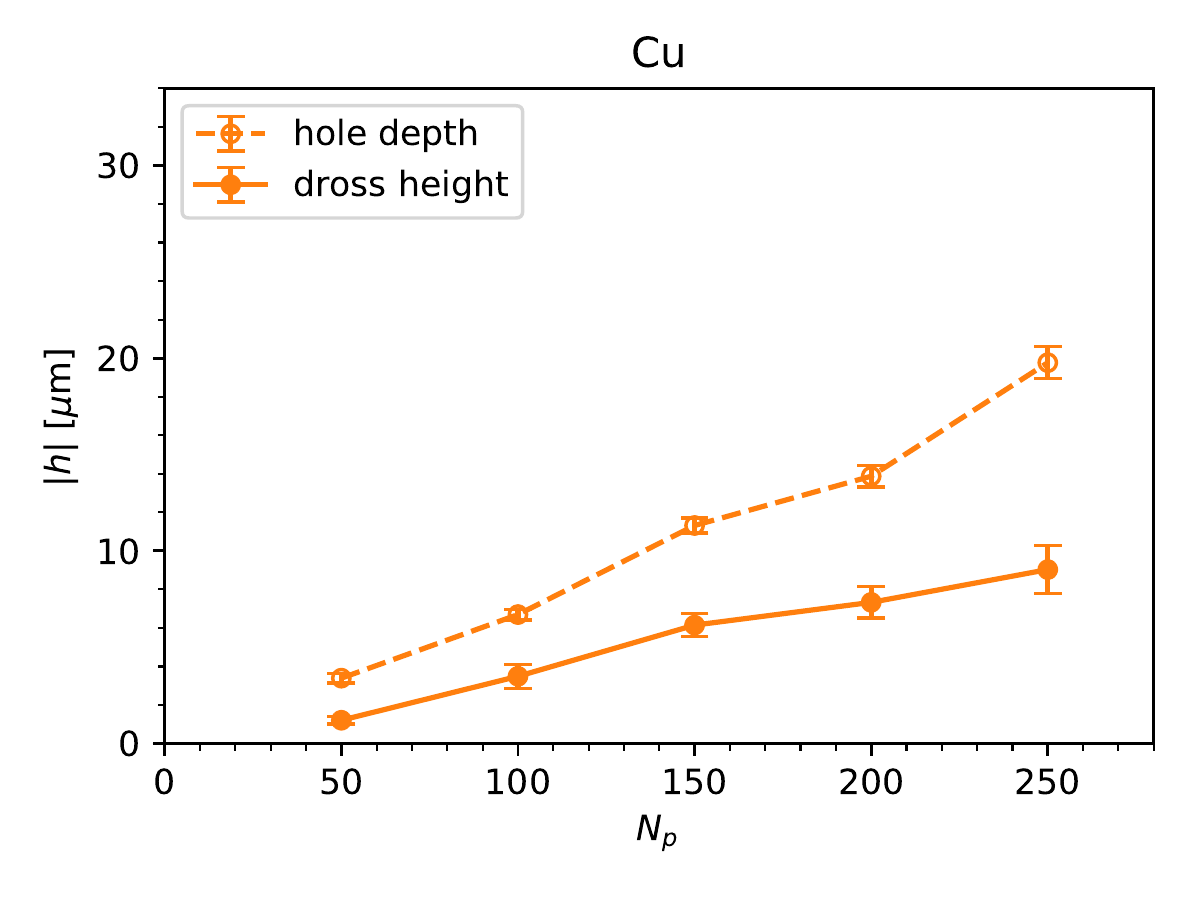}\\
		\includegraphics[width=0.45\linewidth]{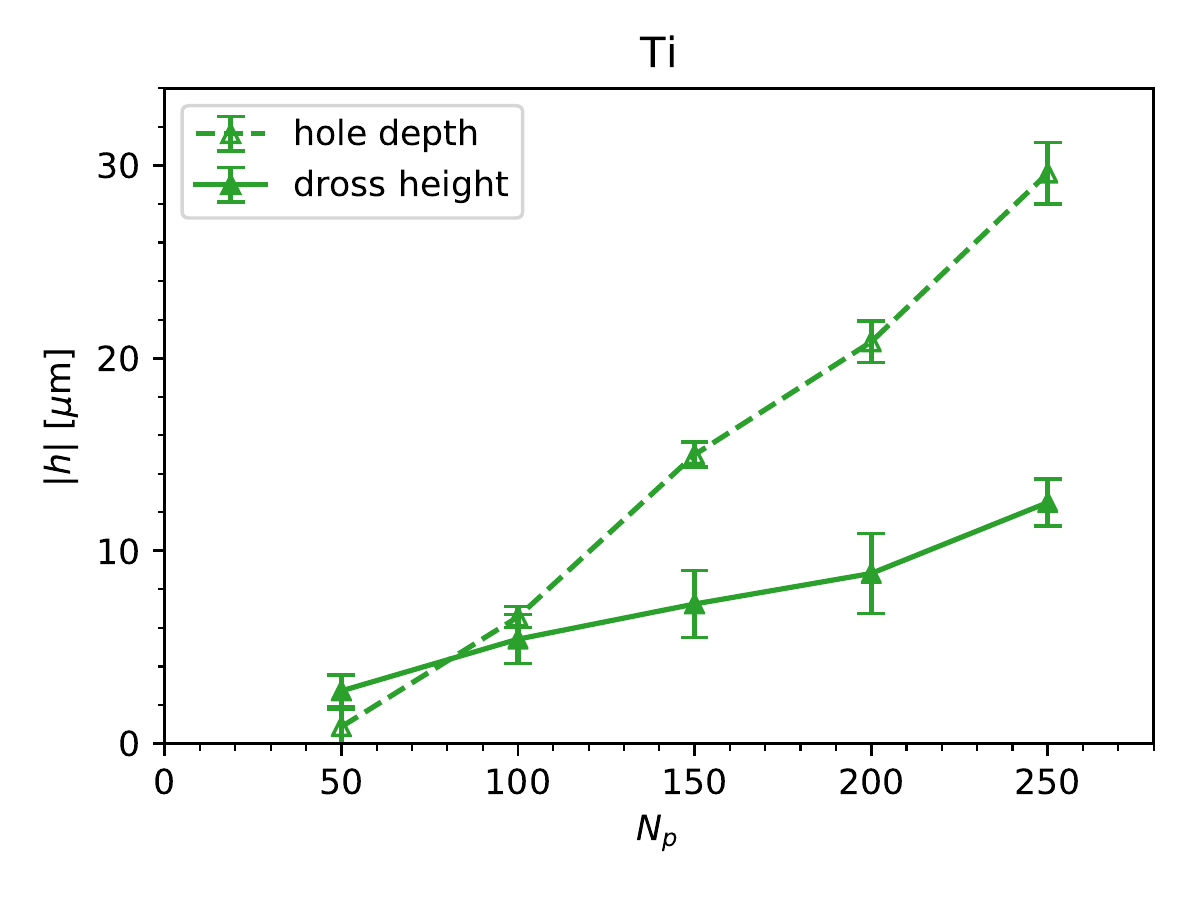}
		\includegraphics[width=0.45\linewidth]{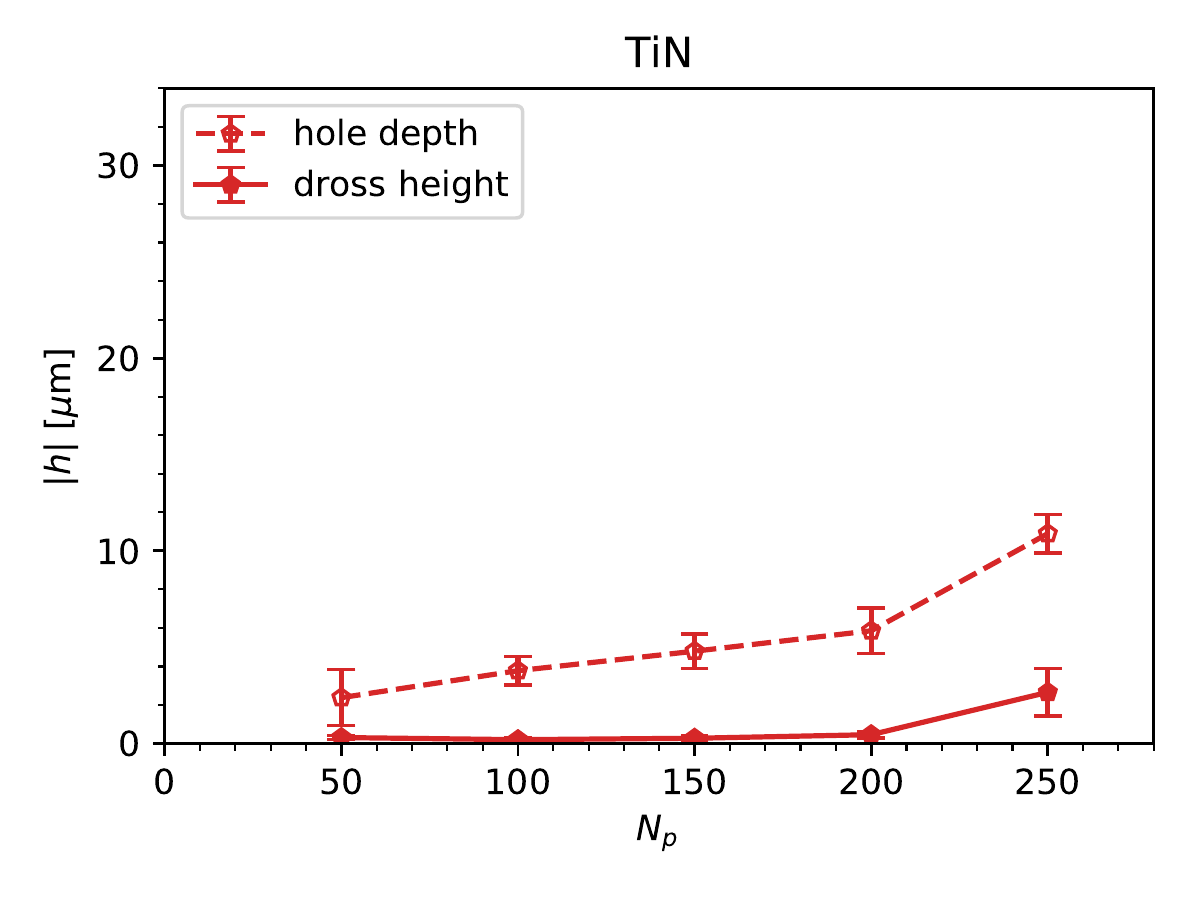}
	\end{center}
	\caption 
	{ \label{fig:hole-depth} Hole depth and dross height as a function of the number of drilling pulses $N_p$ for the different materials.} 
\end{figure}

As a side effect of the ablation process, a fraction of the melt material can solidify around the drilled hole, forming dross and reducing the machining quality, as it can be seen from the illustrations reported in Fig.~\ref{fig:hole-images}. The average hole depth $h_\text{hole}$ and dross height $h_\text{dross}$, measured with 3D microscopy for the samples obtained in the different drilling conditions, are reported in Fig.~\ref{fig:hole-depth}. For all the materials $h_\text{hole}$ and $h_\text{dross}$ tend to increase with the number of pulses $N_p$. TiN exhibits the lowest ablation rate as well as dross, the latter being almost absent.

In order to quantify the average mass ablation rate $M_1$, the height measurements have been fitted to a linear relation. The slope coefficient is used to estimate the single-pulse ablation volume, calculated by assuming, in first approximation, cylindrical holes with a diameter equal to the laser spot size. Then, the ablation mass per pulse $M_1$ has been calculated as the product between the ablated volume and the target density, and the respective values found for each material are reported in Table~\ref{tab:path-fit-coeff}.

Also qualitative differences can be observed between the examined species. In fact, all metallic materials show that the material transport is a mixture of vaporization and melt expulsion, characteristics of the nanosecond-pulsed ablation, with large amounts of dross around the hole aperture. On the other hand, TiN is composed of a granular structure inherent from the PVD process. The focalized thermal field on the surface results in a material transport that is mainly assisted by the disintegration of the coating. Indeed, such a process is slower yet cleaner from a qualitative point of view.

\section{Discussion}
From a qualitative point of view, the model expressed in \eqref{eq:delta-p-t} fits well with the signals of the optical path difference. This can be seen by comparing the simulated behavior of $\Delta p(t)$ in Fig.~\ref{fig:simulation} to the trend of the experimental curves of Fig.~\ref{fig:path-examples}, with the signals increasing and decaying on the expected timescales. However, a quantitative analysis of the results is not straightforward. With the aim of providing just an order of magnitude for the expected value of $\eta$, defined in \eqref{eq:delta-p-eta}, the quantities $K$, $\rho_0$ and $E_0$ can be estimated from the respective typical ranges found in literature and previously introduced while discussing the theoretical model. The effective ablation mass $m_1$ in the plume can be calculated from the estimated average mass $M_1$, reported in Table~\ref{tab:path-fit-coeff}, and from the typical intervals of angular distribution $\theta_p$, as defined in \eqref{eq:mass-distribution}. Following these hypotheses, the order of magnitude for $\eta$ can be estimated within \SI{e-6}{\meter} and \SI{e-9}{\meter}. The values of $\eta$ measured from the maximum optical path difference fall within this wide range. However, a deeper theoretical understanding of the interaction between the SMI measurement and the ablation plume, in combination with extended measurements, would be needed for a precise quantitative comparison.

The analysis of the experimental results for $\Delta p_\text{max}$, presented in Fig.~\ref{fig:max-optical-path-fit}, suggests that the nature of the ablated material influences the amplitude of the interferometric signal, described by the characteristic length $\eta$. Several factors may determine the differences found in the values of $\eta$ reported in Table~\ref{tab:path-fit-coeff}. In the first approximation, a strong dependency of $\eta$ on the energy $E_0$ released in the shock wave can be excluded. In fact $E_0$ is expected to be similar between the different materials, being mainly related to laser pulse energy, which is kept constant.

The differences between materials might be explained in terms of plasma amount and electron number density within the plume \cite{colombo_self-mixing_2017}, as suggested by the relative spectrum intensity increasing with the pulse number. Indeed, the presence of plasma can induce strong changes in the polarization properties of a gas, hence in the effective index of refraction probed by the SMI beam and introduced in \eqref{eq:refractive-index}. However, different physical phenomena might be involved in the visible emission from different materials \cite{margetic_comparison_2000,man_line-broadening_2004}, as suggested by the spectroscopic measurements presented in Fig.~\ref{fig:spectra-materials}. In fact, only few discrete emission lines are visible in the characteristic spectrum of copper, whose weak intensities are comparable to the broad and continuous radiation on the background. Conversely, Ti and TiN spectra are characterized by several strong plasma lines, while stainless steel shows an intermediate behavior. Accordingly, the interferometric measurements, with $\eta$ being similar for the different materials except for TiN, cannot be directly correlated to the spectral characteristics of the radiation emitted during the process, and further interpretations might be needed.

A possible qualitative explanation for the behavior of $\eta$ might come from the morphological analysis of the drilled holes. In fact, as it can be observed from Figures~\ref{fig:hole-images} and~\ref{fig:hole-depth}, for the metallic materials (SST, Cu, Ti) a significant amount of dross is produced from the solidification of melt material, with $h_\text{dross}$ being comparable to $h_\text{hole}$. Conversely, although TiN exhibits a lower absolute ablation rate, the holes obtained for such ceramic material show a negligible formation of dross. It must be noted that the ablated mass that remains in the liquid phase deposited around the hole crater does not contribute to the amount of material within the plume that interacts with the probe beam. As a matter of fact, this may mean that the effective fraction of vaporized mass is higher for TiN, hence justifying its higher value of $\eta$. This suggests that the measure given by the interferometer is strongly influenced by the efficiency of the vaporization mechanism during the microdrilling process, explaining the differences between the examined metallic and ceramic materials.

\section{Conclusion}
This work reported the use of SMI in an inline configuration during laser microdrilling of different metallic and ceramic materials. An analytical model has been developed based on Sedov--Taylor blast wave theory, in order to estimate the characteristic differences between the plume expansion as a function of the pulse number for different processed materials. The main conclusions of the work are summarized as follows:
\begin{itemize}
	\item The SMI beam effectively interacts with the ablation plume, where the signal rise is associated to the change of the refractive index and the expansion of the plume together.
	\item An unwrapping algorithm has been employed, able to resolve the optical path difference below the half-wavelength limit. The behavior of the time-dependent optical path difference could be reconstructed.
	\item The optical path difference depends on the material type as well as on the number of pulses. The unwrapped signals showed a saturating trend toward the end of the process.
	\item Optical emission spectra revealed that the plasma characteristics can contribute to the signal intensity, even if the material expulsion mechanism and quantity are expected to be more relevant.
	\item The maximum optical path difference values were used to estimate the SMI amplitude coefficient. Higher values of this parameter indicate a larger plume extent and a higher refractive index change. The amplitude coefficient of TiN was found to be significantly different compared to the processed metals.
	\item The optical path difference depends highly on the material removal mechanism. A direct transition from solid to vapor phase generates a higher optical path difference. Such process is desirable for improved quality, while the melt phase generation can improve the material removal capacity due to melt expulsion.
\end{itemize}

The results show that SMI can be used to carry out analytical measurements, as well as providing a signal for the process quality monitoring. However, at this level system training is required, where the acceptable signal levels should be determined a priori. The applications can be possibly extended to the measurement of particle flows and contaminants in gases, requiring less stringent temporal and spatial resolution requirements.

\section*{Acknowledgments}
The authors would like to acknowledge Prof. Michele Norgia, Politecnico di Milano, for his contribution in the realization of the self-mixing interferometer setup.

\bibliography{bibliography}

\end{document}